\documentclass[12pt]{article}
\usepackage{graphics}
\usepackage{amssymb}

\begin{document}

\parskip=2pt
\parindent=7mm
\renewcommand{\baselinestretch}{1.}

\newcommand{\be}{\begin{equation}}
\newcommand{\ee}{\end{equation}}
\newcommand{\ba}{\begin{eqnarray}}
\newcommand{\ea}{\end{eqnarray}}
\newcommand{\pd}{\partial}
\newcommand{\CC}{{\mathcal{C}}}
\newcommand{\R}{{\mathbb{R}}}

\title{Berry phase for a potential well transported in a homogeneous
magnetic field}
\date{}
\author{Pavel Exner$^{a,b}\!$ and Vladimir A. Geyler$^{c}\!$}
\maketitle

\begin{quote}
{\small {\em a) Nuclear Physics Institute, Academy of Sciences,
25068 \v Re\v z \\ \phantom{a) }near Prague, Czechia
 \\ b) Doppler Institute, Czech Technical
University, B\v rehov{\'a} 7,\\ \phantom{a) }11519 Prague,
Czechia}
 \\ {\em c) Department of Mathematical Analysis, Mordovian State \\
 \phantom{a) } University,  430000 Saransk, Russia;}\\
 \phantom{a) }\texttt{exner@ujf.cas.cz},
 \texttt{geyler@mrsu.ru} }
\end{quote}

\begin{abstract}
\noindent We consider a two-dimensional particle of charge $e$
interacting with a homogeneous magnetic field perpendicular to the
plane and a potential well which is transported along a closed
loop in the plane. We show that a bound state corresponding to a
simple isolated eigenvalue acquires at that Berry's phase equal to
$2\pi\,{\rm sgn}\,e$ times the number of flux quanta through the
oriented area encircled by the loop. We also argue that this is a
purely quantum effect since the corresponding Hannay angle is
zero.
\end{abstract}


\noindent There are many different situations in which a
nontrivial Berry phase \cite{Ber} is observed. For instance, the
effect was studied recently in mesoscopic systems \cite{LSG,MHK}
for particles with spin interacting with a time-dependent magnetic
field. However, a uniform magnetic field can give rise to a Berry
phase even if the spin-orbital coupling is neglected. An example
has been found in the paper \cite{EG} within a model which
describes a charged particle confined by a point interaction and
placed into a magnetic field of constant direction, which is
independent of time and may be homogeneous; the phase emerges when
the $\delta$ potential (understood in the sense of \cite{AGHH})
moves along a closed loop $\CC$ in the plane.

The aim of this letter is to demonstrate that the same is true
when the point interaction is replaced by any potential capable of
binding the particle, and moreover, that the Berry phase is in
this situation again given explicitly as a multiple of the flux
through the area encircled by $\CC$. The trick we use is adopted
from another example discussed in \cite{EG} being based on the
observation that the transport of the potential well can be
equivalently expressed by means of the magnetic translation group
\cite{Za}.

We consider therefore a two-dimensional spinless quantum particle
described by the Hamiltonian
\be H = H^0 +V(\vec r)\,, \label{Ham} \ee
where
\be H^0 = -\,{\hbar^2\over 2m}\, \Big\lbrack \left(-\pd_x +\pi
i\xi y\right)^2 + \left(-\pd_y -\pi i\xi x\right)^2 \Big\rbrack
\label{Landau} \ee
is the free Schr\"odinger operator with the magnetic field in the
circular gauge,
\be \vec A = {1\over 2}\, \vec B \times \vec r\,, \ee
where $\vec B= B\vec e_3$ is the field strength and the quantity
$\xi$ appearing in (\ref{Landau}) is the magnetic flux density,
\be \xi = {B\over\Phi_0}\, \mathrm{sgn}\,e \ee
with $\Phi_0= 2\pi\hbar c/|e|$ being the flux quantum. For the
sake of simplicity we employ in the following the rational system
of units, $|e|=\hbar=c=2m=1$.

We suppose that the potential $V$ is such that the operator
(\ref{Ham}) is self-adjoint and has at least one simple eigenvalue
which we call $E_0$; the corresponding normalized eigenfunction
will be denoted as $\psi_0$. We will consider the family of
shifted operators
\be H(\vec a) = H^0 +V(\vec r-\vec a) \label{Ham-a} \ee
and denote by $V_{\vec a}$ the operator of multiplication by
$V(\vec r-\vec a)$. Let $[\vec a]$ be the operator of magnetic
translation on the vector $\vec a$,
\be [\vec a]f(\vec r) = \exp(-\pi i\xi \vec r\wedge \vec a)\,
f(\vec r-\vec a)\,, \label{mgtrans} \ee
then the intertwining relation
\be [\vec a]V_0 = V_{\vec a}[\vec a] \ee
is valid. It follows that $\psi_{\vec a}(\vec r) = [\vec a]
\psi_0(\vec r)$ is an eigenfunction of the operator $H(\vec a)$
corresponding to the eigenvalue $E_0$.

We want to find the Berry phase which refers to an adiabatic
evolution of the system with the Hamiltonian $H(\vec a)$ when the
vector $\vec a$ which characterizes the potential position moves
along a loop $\CC$. To this end we have to find the corresponding
Berry potential
\be U(\vec a) = i\langle \psi_{\vec a}| \nabla_{\vec a} \psi_{\vec
a} \rangle \,. \ee
From (\ref{mgtrans}) we get
\ba \pd_{a_1}\psi_{\vec a}(x,y) &\!=\!& \pi i\xi y \psi_{\vec
a}(x,y) + \exp(-\pi i\xi \vec r\wedge \vec a)\, \pd_{a_1}
\psi_0(x\!-\!a_1, y\!-\!a_2) \nonumber \\ &\!=\!& \pi i\xi y
\psi_{\vec a}(x,y) - \exp(-\pi i\xi \vec r\wedge \vec a)\, \pd_x
\psi_0(x\!-\!a_1, y\!-\!a_2) \ea
and
$$ \pd_{a_2}\psi_{\vec a}(x,y) = -\pi i\xi x \psi_{\vec a}(x,y) -
\exp(-\pi i\xi \vec r\wedge \vec a)\, \pd_y \psi_0(x\!-\!a_1,
y\!-\!a_2)\,. $$
Consequently,
\ba \langle \psi_{\vec a}| \pd_{a_1}\psi_{\vec a}\rangle &\!=\!&
\int \!\!\int \pi i\xi (y\!-\!a_2) |\psi_0(x\!-\!a_1,y\!-\!a_2)|^2
dx\,dy \nonumber \\ &\!+\!& \pi i\xi a_2 \int \!\!\int
|\psi_0(x,y)|^2 dx\,dy \nonumber \\ &\!-\!& \int \!\!\int
\overline{\psi_0(x\!-\!a_1, y\!-\!a_2)}\, \pd_x \psi_0(x\!-\!a_1,
y\!-\!a_2)\, dx\,dy \nonumber \\ &\!=\!& \pi i\xi a_2 + \pi i\xi
c_1\,, \ea
where
\be c_1 = \int \!\!\int y\, |\psi_0(x,y)|^2 dx\,dy - {1\over \pi
i\xi}\, \int \!\!\int \overline{\psi_0(x,y)}\, \pd_x \psi_0(x,y)\,
dx\,dy \ee
is independent of $\vec a$. Analogously,
\be \langle \psi_{\vec a}| \pd_{a_2}\psi_{\vec a}\rangle = -\pi
i\xi a_1 + \pi i\xi c_2 \ee
with
\be c_2 = \int \!\!\int x\, |\psi_0(x,y)|^2 dx\,dy - {1\over \pi
i\xi}\, \int \!\!\int \overline{\psi_0(x,y)}\, \pd_y \psi_0(x,y)\,
dx\,dy\,.  \ee
This yields
$$ U_1(\vec a) = -\pi \xi a_2 - \pi \xi c_1\,, \quad U_2(\vec a) =
\pi \xi a_1 - \pi \xi c_2\,. $$
Moreover, the second terms can be removed by an appropriate gauge
transformation, so we get
\be U(\vec a) = \pi \xi (-a_2,a_1)\,; \label{Bpot} \ee
notice that the expression coincides with the vector potential of
the homogeneous magnetic field. Using (\ref{Bpot}) we find that
the corresponding Berry phase is given by
\be \gamma(\CC) = \pi \xi \int_{\CC} (a_1 da_2 - a_2da_1) = 2\pi
\xi S\,, \ee
where $S$ is the (oriented) area encircled by the loop $\CC$, in
other words,
\be \gamma(\CC) = 2\pi\, \mathrm{sgn}\,e\, {\Phi_{\CC}\over
\Phi_0}\,, \ee
where $\Phi_{\CC}/\Phi_0$ is the number of flux quanta through
$S$. This is the result we have announced in the opening.

Let us add a comment on the meaning of the result. Consider a
classical dynamical system with the Hamiltonian $H^0+V(\vec r)$
which is completely integrable. Recall that the Hannay angle
$\Delta\theta(\CC)$ can be defined \cite{Han} for a cyclic
adiabatic evolution of the system with the ``shifted'' Hamiltonian
$H(\vec a)= H^0+V(\vec r\!-\!\vec a)$ along a closed loop $\CC$ is
the parameter space $\R^2$; this angle is a classical counterpart
to the Berry phase \cite{Ber2}. In our case, however, we have
$\Delta\theta(\CC)=0$ for any loop $\CC$. Indeed, since parallel
translation are canonical transformation on the phase space --
which follows from the fact that $H^0$ is a quadratic polynomial
in coordinate and momentum coordinates -- all the Hamiltonians
$H(\vec a)$ have identical canonical frequencies. If, moreover,
the potential is such that $H(\vec 0)$ has a purely discrete
spectrum, then the vanishing of the Hannay angle is a consequence
of the Berry correspondence principle \cite{Ber2},
$\Delta\theta(\CC) = -\pd \gamma_n(\CC)/\pd n$, since in our case
the Berry phase is independent of the index $n$ labeling the
energy levels.

The appearance of a nontrivial adiabatic phase factor in a moving
potential $V_{\vec a}$ is thus a purely quantum phenomenon like,
e.g., the well-known Aharonov-Bohm effect. This feature
distinguished the effect discussed in this letter from the Berry
phase related to the Larmor precession of an electron bound to a
fixed centre of attraction such as a heavy nucleus \cite{Ham}.

Let us finish with another remark concerning an extension of the
above result to three-dimensional systems. Since the homogeneous
field is parallel to the $z$-axis we find easily that $U_3(\vec
a)=0$. Consequently, the Berry phase along a closed loop $\CC$ is
again $\gamma(\CC)=2\pi\, \mathrm{sgn}\,e\,\Phi_C/\Phi_0$, up to a
sign, where $\Phi_C$ is now the magnetic flux through the
projection of $\CC$ to a plane perpendicular to the field.


\subsection*{Acknowledgment}

The research has been partially supported by GAAS and Czech
Ministry of Education under the contracts 1048801 and ME099.


\end{document}